\documentclass[aps,pre,showpacs,superscriptaddress]{revtex4}
\usepackage[dvips]{graphicx}
\usepackage{subfigure}
\begin{document}

\title{Theory of Hantavirus Infection Spread Incorporating Localized
Adult and
Itinerant Juvenile Mice}

\author{V. M. Kenkre}
\affiliation{Consortium of the Americas for Interdisciplinary Science
and
Department of Physics and Astronomy,
University of New Mexico, Albuquerque, New Mexico 87131, USA}

\author{L. Giuggioli}
\affiliation{Consortium of the Americas for Interdisciplinary Science
and
Department of Physics and Astronomy,
University of New Mexico, Albuquerque, New Mexico 87131, USA}

\author{G. Abramson}
\affiliation{Consortium of the Americas for Interdisciplinary Science
and
Department of Physics and Astronomy,
University of New Mexico, Albuquerque, New Mexico 87131, USA}
\affiliation{Centro At\'{o}mico Bariloche, CONICET and Instituto
Balseiro,
8400 San Carlos de Bariloche, R\'{\i}o Negro, Argentina.}

\author{G. Camelo-Neto}
\affiliation{Consortium of the Americas for Interdisciplinary Science
and
Department of Physics and Astronomy,
University of New Mexico, Albuquerque, New Mexico 87131, USA}

\date{\today}

\begin{abstract}
A generalized model of the spread of the Hantavirus in mice populations
is
presented on the basis of recent observational findings concerning the
movement characteristics of the mice that carry the infection. The
factual
information behind the generalization is based on mark-recapture
observations reported in Giuggioli et al. [Bull. Math. Biol. \textbf{67}
1135 (2005)] that have
necessitated the introduction of home ranges in the simple model of
Hantavirus spread presented
by Abramson and Kenkre [Phys. Rev. E \textbf{66} 11912 (2002)]. The
essential feature of the model presented here is the existence of adult
mice
that remain largely confined to locations near their home ranges, and
itinerant juvenile mice that are not so confined, and, during their
search
for their own homes, move and infect both other juveniles and adults
that
they meet during their movement. The model is presented at three levels
of description: mean field, kinetic and configuration. Results of
calculations are shown explicitly from the mean field equations and the
simulation rules, and are found to agree in some respects and to differ
in others. The origin of the differences is shown to lie in spatial
correlations. It is indicated how mark-recapture observations in the field may be employed to verify the applicability of the theory.
\end{abstract}

\maketitle

\section{Background and Motivation for the Study}

\label{sec:background}

The spread of epidemics is an important topic that has received a great
deal
of attention from researchers in recent times \cite
{yatesreview,mills99,ak0,akyptraveling,vmkpasi,vmkphysica}. This
interest
stems from multiple factors. On the utilitarian side, concerns regarding
human health provide an obvious reason for carrying out such studies. An
equally important motivation arises purely from intellectual sources:
the
desire to gain a general understanding of spatially resolved strongly
interacting systems on a macroscopic scale. Examples of epidemics of
particular interest to the interdisciplinary scientist are the plague,
the
West Nile virus, and the Hantavirus. From the many-body aspect in
(theoretical) modeling activities, the first presents formidable
problems
related to the number of carriers and diverse interactions of the fleas
and
their connections to their hosts. The second provides an interesting
example
of a system with two carriers, mosquitoes and birds, with disparate
lifespans (weeks and years respectively) and suggests \cite{westnile}
time scale disparity
analyses known in condensed matter sciences (see e.g. a discussion of
the
Born-Oppenheimer approximation in solids in ref. \cite{wnm}). The
simplest
of the epidemics to study, from the conceptual viewpoint of a theorist,
is
the Hantavirus \cite
{yatesreview,mills99,ak0,akyptraveling,vmkpasi,vmkphysica}. It is the
subject of the study presented below.

The details of the Hantavirus epidemic may be found in the review by
Yates et al. \cite{yatesreview} and related refs. \cite
{mills99,ak0,akyptraveling,vmkpasi,vmkphysica}. The essential
feature of the epidemic from the modeling point of view is that the
infection is carried by mice that move from one physical location to
another, and is transmitted to other mice through what are probably
aggressive encounters. It is believed that the mice and the virus have
coexisted for millions of years and therefore the mice do not die, nor
are
otherwise impaired, from contraction of the virus. This feature is
peculiar
to the Hantavirus and not shared with other epidemics such as the
plague or
the West Nile virus where the carriers may die from contracting the
virus.
Oscillations in the population which are characteristic of such
behavior are
therefore absent in the Hantavirus. Furthermore, in the Hantavirus
context,
no mice are born infected: infection may only be contracted from other
mice
after birth: there is no ``vertical transmission'' of the disease. The
human
population is incidental to the evolution of infection within the mouse
population since humans get the virus from the mice but have no feedback
effects on the mice in the infection process.

These features of the epidemic led a few years ago to the construction
by
two of the present authors of a simple model of the spread of the
Hantavirus
in the mouse population. In addition to the above features particular
to the
epidemic, interaction of the mice with the environment through standard
logistic terms \cite{murray,okubobook} and motion of the mice over the
terrain
viewed as a simple and free random walk were the elements that went
into the
making of the model. The model was introduced at the usual three levels
of
the description of a many-body system: the most crude, the mean field
level,
the more detailed, the kinetic level, and the most detailed, the
configuration level. The last was handled through simulations
\cite{aguirre}%
. The first two levels were treated in ref. \cite{ak0}, and most active
studies were carried out at the kinetic level \cite{ak0} which describes
through the use of partial differential equations, the time evolution
of the
average but spatially resolved mice populations $M_s$ (susceptible) and
$M_i$
(infected). The specific equations, with $t$ denoting time and $x$
representing location in a space of appropriate dimensions, a 2-d
description being typically sufficient, are
\begin{eqnarray}
\frac{\partial M_s}{\partial t} &=& b (M_s+M_i)-cM_s-\frac{M_s
(M_s+M_i)}{%
K(x,t)}-aM_s M_i+D\nabla^{2} M_s,  \nonumber  \label{ak1} \\
\frac{\partial M_i}{\partial t} &=& -cM_i-\frac{M_i
(M_s+M_i)}{K(x,t)}+aM_s
M_i+D\nabla^{2} M_i.  \label{akmodel}
\end{eqnarray}
Considerations such as those arising from gender and age of the animals
were
tacitly neglected, and all parameters except $K$ were considered to be
independent of time.

The parameters of the model are $a,b,c,K$ and $D.$ All processes are
assumed
to be occurring continuously in time, an approximation which seems
reasonable unless observations or phenomena are particularly chosen to
be probed
at very short time scales. The processes of birth and death of the mice
are
represented as occurring at rates $b$ and $c$, respectively. The
transmission
of infection occurs through encounters, the aggression parameter being
$a$.
Saturation of the population is assumed to occur by competition for the
resources among the mice through the environmental parameter $K$ which
describes nutrition available to the mice. This parameter, whose product
with $b-c$ is called the `carrying capacity', controls the so-called
logistic term which resolves in the standard manner the paradox of
Malthusian explosion of the population.

Let us focus on this kinetic level model and refer to it as has been
done in
the current literature \cite{aguirre,k1} as the AK model. It may be
regarded
from the ecological viewpoint merely as the familiar SI model extended
to
include spatial resolution and diffusive transport. From the
mathematical
point of view it may be said to represent a system obeying the Fisher
equation \cite{murray} with \emph{internal states} representing
infection or
its absence, respectively. While near-trivial to conceive, this model
has
had considerable success in the short time since it was proposed for the
Hantavirus \cite{ak0}. It has led to qualitative and semi-quantitative
success in explaining observations such as spatio-temporal patterns in
the
epidemics. These patterns are associated with correlations between
periods
of precipitation and epidemic outbreaks, and with the spatial location
of
refugia--regions of the landscape in which infection persists during
off-periods of the epidemic \cite{ak0,vmkpasi}. Other applications of
the
model include the detailed understanding and control of traveling waves
of
infection \cite{akyptraveling}, fluctuations arising from the
finiteness of
the numbers and discreteness of the population of the rodents \cite
{aguirre,k1}, environmental effects \cite{bkk}, curious switching
effects
that have been predicted to occur \cite{k2}, and extensions to unrelated
systems such as bacteria in Petri dishes \cite{lucaadv,kkmask}.

The success of and interest in this so-called AK model represented by
(\ref
{akmodel}) led to recent attempts to devise practical prescriptions for
the
extraction of the parameters constituting the model from field
measurements.
The mouse birth rate $b$ and the death rate $c$ are obtained from field
observations without too much trouble and are generally considered to be
constants in space and time. With some effort, reasonable estimates of
the
environment resource parameter $K\left( x,t\right) $ as a function of
location and time can be obtained by counting food (such as nuts and
water)
available to the mice in the different locations, as well as by
acquiring
aerial photographs of the vegetation cover. Relative, rather than
absolute,
quantification of $K$ is possible in this way. Observational collection
of
data concerning the aggression rate $a$, through which infection is
thought
to be transmitted during mouse-mouse encounters, turns out to be so
difficult that, at least at the present moment, it must be considered an
adjustable parameter. It was, however, possible to focus on the
important
parameter, the mouse diffusion constant $D$, and to obtain it
quantitatively
from field measurements \cite{panama,sevilleta,jtb}.

The basic idea behind the extraction of $D$ was to regard to the extent
possible that the mouse movement is a simple random walk, and to
extract $D$
from records of the movement through the use of the well-known
proportionality of the mean square displacement to $Dt$. The details
of the theory \cite{jtb} and the implementation of the prescriptions
obtained from the theory to mark-recapture observations carried out in
Panama \cite{panama} and New Mexico \cite{sevilleta} may be found in our
recent work. The important and perhaps surprising conclusion that
emerged
from that work was that the mouse mean square displacement, which grows
linearly with $t$ for short times, is found to \emph{saturate} at large
times. The appearance of a length scale in the random movements of the
mice
may be ascribed to the fact that animals typically move near fixed
locations
(burrows) for reasons of shelter and security \cite
{parmenter83,stickel,wolff97}, but it could also be ascribed to the fact
that mark-recapture observations employ a limited region of space where
the
traps are laid out. It is possible to show analytically \cite{panama}
that
either of these factors could independently lead to the saturation of
the
mean square displacement. A disentangling of the two length scales was
possible and led to explicit deductions of both the diffusion constant
of
the mice and their home ranges. Thus, reasonable realistic extracted
values
of the home range size $L$ for different types of mice in different
environments were found to be between 50 and 120 m for
\emph{Zygodontomys brevicauda}
in Panama and about 100 $\pm$ 25 m for \emph{Peromyscus maniculatus} in
New Mexico.
The respective values of the diffusion constant $D$ of the mice turned
out
to be 200 $\pm$ 50 and 470 $\pm$ 50 meters squared per day.

This quantitative information has provided the impetus (indeed,
necessity)
to generalize the AK model expressed in Eq. (\ref{akmodel}) to
incorporate home
ranges. Kenkre has suggested \cite{vmkphysica05} several different
generalizations for this purpose. One simple way of incorporating home
ranges in our model of epidemic spread is to add potential terms to Eq.
  (\ref
{akmodel}). Such an analysis, carried out by MacInnis et al.
\cite{david} has
resulted in modifications in the AK predictions for refugia sizes
and shapes. Another, simpler, model modification in the AK equations
has led
us \cite{kg} to apply the so-called Montroll defect technique
\cite{Montrolldefect}
to deduce memory-possessing variations of the AK equations on the one
hand and
time-dependent diffusion constant variations on the other. Perhaps the
most
fertile model that has emerged from the work on the determination of
motion
and demographic characteristics is the one that we discuss in the
present
paper. It is set out in Sec. \ref{sec:model} below and an examination of its
consequences form the rest of the paper.

\section{A generalized model for the spread of the Hantavirus}

\label{sec:model}

Let us consider the dynamics of two types of mice, stationary and
itinerant
(and susceptible and infected in each category). The stationary mice
are the
adults that move within their home ranges and do not stray far from the
burrow. We have termed them `localized adults' in the title of the
present
paper. For the sake of simplicity as well as to emphasize the new
features
that appear as a result of confinement, we neglect here the fact that
home
ranges may overlap and lead thereby to transmission of infection. The
itinerant mice are the subadults (called `itinerant juveniles' in the
title)
that must leave to find their own home ranges. Adults do not move
because
their home ranges are considered of negligible extent for the purposes
of
this description. They die at the rate $c$ and have the standard
logistic
competition interactions with the environment controlled by the
environment
parameter $K$. They may be infected or not, the only possibility of
their
contracting infection being when an infected juvenile visits their home
range. If infected, they may transmit infection to a susceptible
juvenile if
it visits their home range. Adults are not born but juveniles turn into
adults. This happens when a juvenile finds an appropriate site to settle
down in, which then becomes its burrow.

Juveniles are born at a rate $b$ from the adult population. They are
mobile,
their motion being diffusive. They may acquire or transmit infection to
other adults or other juveniles on encounter. If they find an
appropriate
site they turn into adults and become immobile as described above. They
also
have environment competition rates with the rest of the mice. In order
to
focus attention on special features of our generalization, we neglect
the
death rate $c_B$ of the juveniles and allow their population to be
depleted
only through the competition term and their conversion to adults through
growth. In a companion paper \cite{ana} we have included the death rate
of
the juveniles and indeed examined the specific effect of changes in that
rate induced by the existence of predators in the open field.

The characteristics that we have described above suggest that the AK
model
in (\ref{akmodel}) be replaced by
\begin{eqnarray}
\frac{\partial B_{i}(x,t)}{\partial t}
&=&-c_{B}B_{i}-\frac{B_{i}(A+B)}{%
K(x,t)}+aB_{s}(A_{i}+B_{i})+D\nabla ^{2}B_{i}-G(x) B_{i},  \nonumber
\\
\frac{\partial B_{s}(x,t)}{\partial t}
&=&bA-c_{B}B_{s}-\frac{B_{s}(A+B)}{%
K(x,t)}-aB_{s}(A_{i}+B_{i})+D\nabla ^{2}B_{s}-G(x) B_{s},  \nonumber
\\
\frac{\partial A_{i}(x,t)}{\partial t} &=&-cA_{i}-\frac{A_{i}(A+B)}{%
K(x,t)}+aA_{s}B_{i}+G(x) B_{i},\nonumber \\
\frac{\partial A_{s}(x,t)}{\partial t} &=&-cA_{s}-\frac{A_{s}(A+B)}{%
K(x,t)}-aA_{s}B_{i}+G(x) B_{s}.
\label{lsmodel}
\end{eqnarray}
where $A$ and $B$ (without suffixes) denote the total densities of the
adult
and juvenile mice respectively, the suffixes $i$ and $s$ represent
infected
and susceptible states as earlier, and the last terms in each equation
describe the
settling down of the juveniles into their own homes, accompanied by
their
conversion into (static) adults. The rate of such conversion is $G(x)$.
This $G(x)$ is non-zero only if $x$ lies in the `green
pastures', that is the spatial regions that the juveniles find suitable
as
their home ranges. Note that there are no spatial derivatives in the
equations for the adults because the adults do not move, that being an
extreme
representation of their confinement to their home ranges.

The model represented in Eq. (\ref{lsmodel}) describes the processes at
a
kinetic level as does the AK model in Eq. (\ref{akmodel}). This means
that the
key quantities are mice \emph{densities} and that the evolution is
described
via \emph{kinetic} equations such as the Fisher equation \cite{okubobook,murray}. A less
detailed
description, with no spatial resolution included, is provided by a mean
field model in which the key quantities are the total mice numbers $\int
A(x,t)dx$ and $\int B(x,t)dx$ where the integrals are over the entire
landscape. A \emph{more} detailed description than in kinetic models is
provided by configuration master equation approaches (see for example
\cite
{kac,kenkrevanhorn}) which include fluctuation effects. In the present
paper we
will analyze both the mean field and the configuration master equation
approaches, reserving the middle-level kinetic treatment for a future
publication.

Our paper is laid out as follows. In Sec. \ref{sec:analysis}, we
present
an explicit analysis of the mean field model, and the simulation
treatment
of the configuration level approach. In Sec. \ref{results}, we
present
essential results and their discussion. In Sec. \ref{conclusions}, we
present concluding remarks.

\section{Analysis: mean field and simulation treatments}

\label{sec:analysis}

In this section we treat the model of Eq. (\ref{lsmodel}) first by
simplifying
it to the mean field level of description and then by augmenting it to
the
configuration level description. For simplicity we keep $K$ time and
space
independent.

\subsection{Mean field Description}

\label{sec:meanfield}

The mean field description focuses on the time evolution of the
integrals of
the densities in a kinetic description such as that of Eq.
(\ref{lsmodel})
and thereby loses space resolution. Called in some contexts the
`well-stirred
limit', the mean field description may be considered as the limit of
Eq. (%
\ref{lsmodel}) for an infinitely large diffusion constant $D$.
In passing from the kinetic to the mean field description,
the single infection quantity $a$ in Eq. (\ref{lsmodel})
results in \emph{two} corresponding quantities: $a_{0}$ and $a_{1}$.
The former (latter) refers to the transfer of infection between an
adult (juvenile)
and a juvenile. Both quantities result from the combination of the
infection event
and the motion process. The latter is explicit in Eq. (\ref{lsmodel})
but not in
Eq. (\ref{normset}). The difference between $a_{0}$ and $a_{1}$ is
precisely
the difference between the expression $4\pi RD$ and $8\pi RD$ which
describes
the capture and mutual annihilation rates respectively in the
literature on
excitons in molecular crystals \cite{KenkreGMEbook,PopeSwenbergbook}.
Although the precise relationship of $a_{0}$ and $a_{1}$ would depend
on the relative
importance of the mice motion and infection processes, we will take
$a_{1}=2a_{0}$
for simplicity in the rest of the paper in keeping with the extreme
limits
considered in other literature contexts \cite{Smoluchowski}.

The juveniles transmit infection among themselves with rate per unit
density
$a_{1}$, get infected and transmit infection to the adults with rate per
unit density $a_{0}$, struggle for resources all over space through the
environment parameter $K$ and become adult with a growth rate $g$. This
growth rate, introduced to represent  a juvenile settling into an
unoccupied home range and growing into an adult, is proportional to
$G(x)$ of Eq. (\ref{lsmodel}). The other
terms can be interpreted as explained above for Eq. (\ref{lsmodel}). The
coupled set of equations for normalized quantities in the mean field
description is
\begin{eqnarray}
\frac{d\mathcal{B}_{i}}{d\tau } &=&-\gamma \mathcal{B}_{i}+\alpha _{0}%
\mathcal{B}_{s}\mathcal{A}_{i}+\alpha
_{1}\mathcal{B}_{i}\mathcal{B}_{s}-%
\mathcal{B}_{i}\left( \mathcal{A}+\mathcal{B}\right) ,  \nonumber \\
\frac{d\mathcal{B}_{s}}{d\tau } &=&-\gamma \mathcal{B}_{s}+\beta \left(
\mathcal{A}_{s}+\mathcal{A}_{i}\right) -\alpha
_{0}\mathcal{B}_{s}\mathcal{A}%
_{i}-\alpha _{1}\mathcal{B}_{i}\mathcal{B}_{s}-\mathcal{B}_{s}\left(
\mathcal{A}+\mathcal{B}\right),  \nonumber \\
\frac{d\mathcal{A}_{i}}{d\tau } &=&-\mathcal{A}_{i}+\alpha
_{0}\mathcal{A}%
_{s}\mathcal{B}_{i}+\gamma \mathcal{B}_{i}-\mathcal{A}_{i}\left(
\mathcal{A}+%
\mathcal{B}\right) ,  \nonumber \\
\frac{d\mathcal{A}_{s}}{d\tau } &=&-\mathcal{A}_{s}-\alpha
_{0}\mathcal{A}%
_{s}\mathcal{B}_{i}+\gamma \mathcal{B}_{s}-\mathcal{A}_{s}\left(
\mathcal{A}+%
\mathcal{B}\right).
\label{normset}
\end{eqnarray}
where each script character denotes the ratio of the quantity described
by the
corresponding Roman character and $cK$. Thus,
$\mathcal{A}_{i}=A_{i}/(cK)$. We also write $\tau =ct$ and the
dimensionless
parameters are now $\gamma =g/c$, $\beta =b/c$, $\alpha _{0}=Ka_{0}$%
, and $\alpha _{1}=Ka_{1}$. As explained above we take $a_{1}=2a_{0}$.

From Eq. (\ref{normset}) it is easy to recognize that the total adult
and
juvenile populations, respectively,
$\mathcal{A}=\mathcal{A}_{i}+\mathcal{A}%
_{s}$ and $\mathcal{B}=\mathcal{B}_{i}+\mathcal{B}_{s}$, obey the
following
evolution
\begin{eqnarray}
\frac{d\mathcal{B}}{d\tau } &=&\beta \mathcal{A}-\gamma
\mathcal{B}-\mathcal{%
B}(\mathcal{A}+\mathcal{B}) ,  \nonumber \\
\frac{d\mathcal{A}}{d\tau } &=&-\mathcal{A}+\gamma
\mathcal{B}-\mathcal{A}(%
\mathcal{A}+\mathcal{B}).  \label{sumevol}
\end{eqnarray}
Even if $\mathcal{A}$ and $\mathcal{B}$ are both limited by a quadratic
saturation, it is evident from (\ref{sumevol}) that the sum
$\mathcal{A}+%
\mathcal{B}$ does not obey the standard logistic equation. This fact,
as we will see in Sec.
\ref{results}, gives some qualitative differences of the steady state
parameters dependence of (\ref{normset}) from the AK model. The steady
state values
$\overline{\mathcal{A}}$ and $\overline{\mathcal{B}}$ are given  for
$\beta >1$  by
\begin{eqnarray}
\overline{\mathcal{A}} &=&\gamma \frac{1+\xi -\sqrt{1+2\xi }}{\xi },
\nonumber \\
\overline{\mathcal{B}} &=&\frac{1+\xi -\sqrt{1+2\xi }}{\xi }\left[
1+\frac{%
(\gamma +1)}{2}\left( \sqrt{1+2\xi }-1\right) \right] ,  \label{sumss}
\end{eqnarray}
with $\xi =2(\beta -1)\gamma /(1+\gamma )^{2}$. The situation $\beta>1$
represents the juvenile birth rate $b$ being larger than the
adult death rate $c$. For the opposite situation, $\beta <1$, when the
adults die quicker than the juveniles are born, the
trivial solution $\overline{\mathcal{A}}=\overline{\mathcal{B}}=0$
emerges
from Eq. (\ref{sumevol}). These two steady states exchange their
stability as $%
\beta $ crosses the value $1$ clearly indicating the presence of a
transcritical bifurcation at $\beta =1$ from zero to non zero population
density.
\begin{figure}[ht]
\centering
\mbox{\subfigure{\label{Fig1a}\includegraphics[height=0.4\columnwidth,width=0.46\columnwidth]{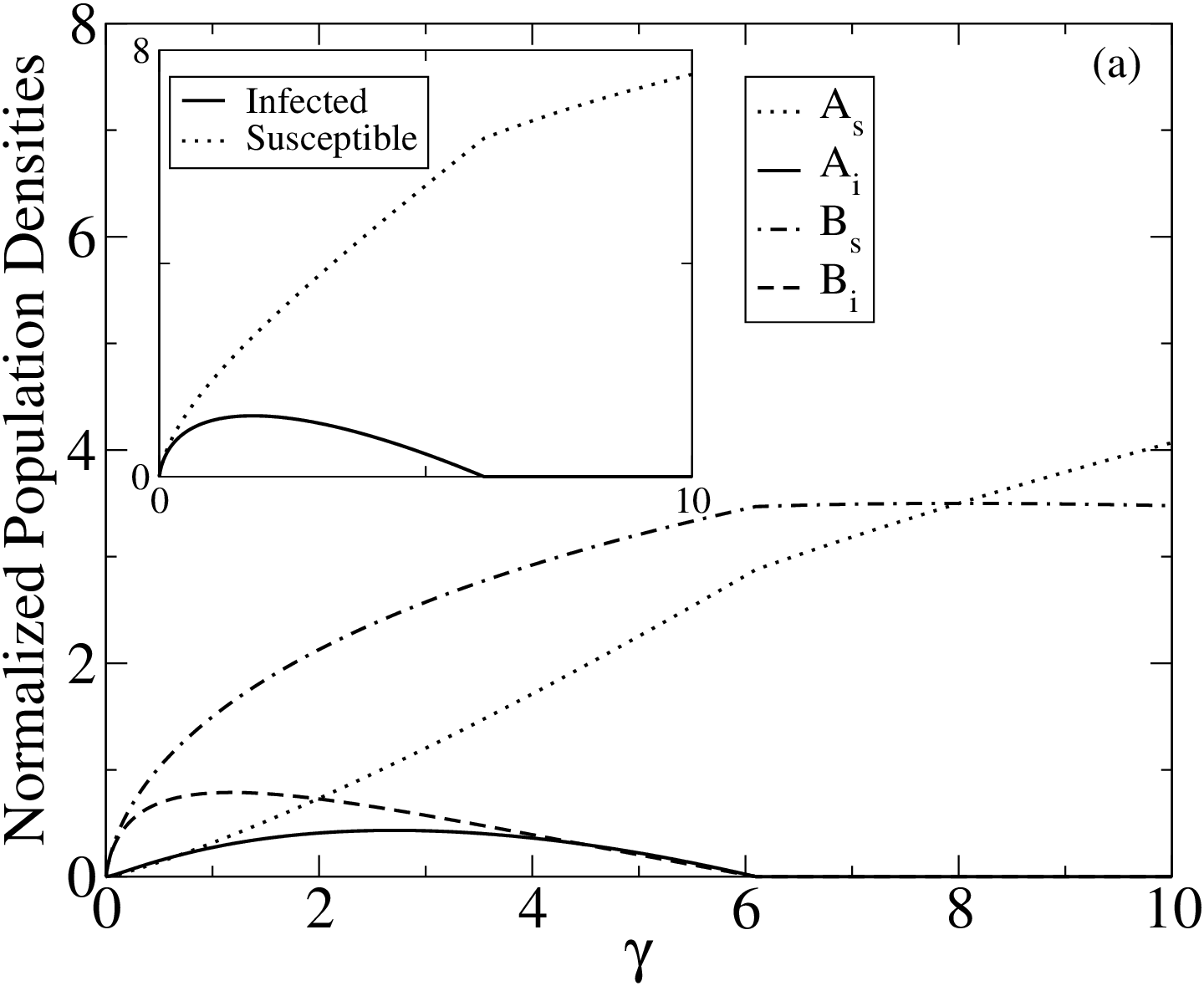}}
\hspace{0.03\columnwidth}
\subfigure{\label{Fig1b}\includegraphics[height=0.4\columnwidth,width=0.46\columnwidth]{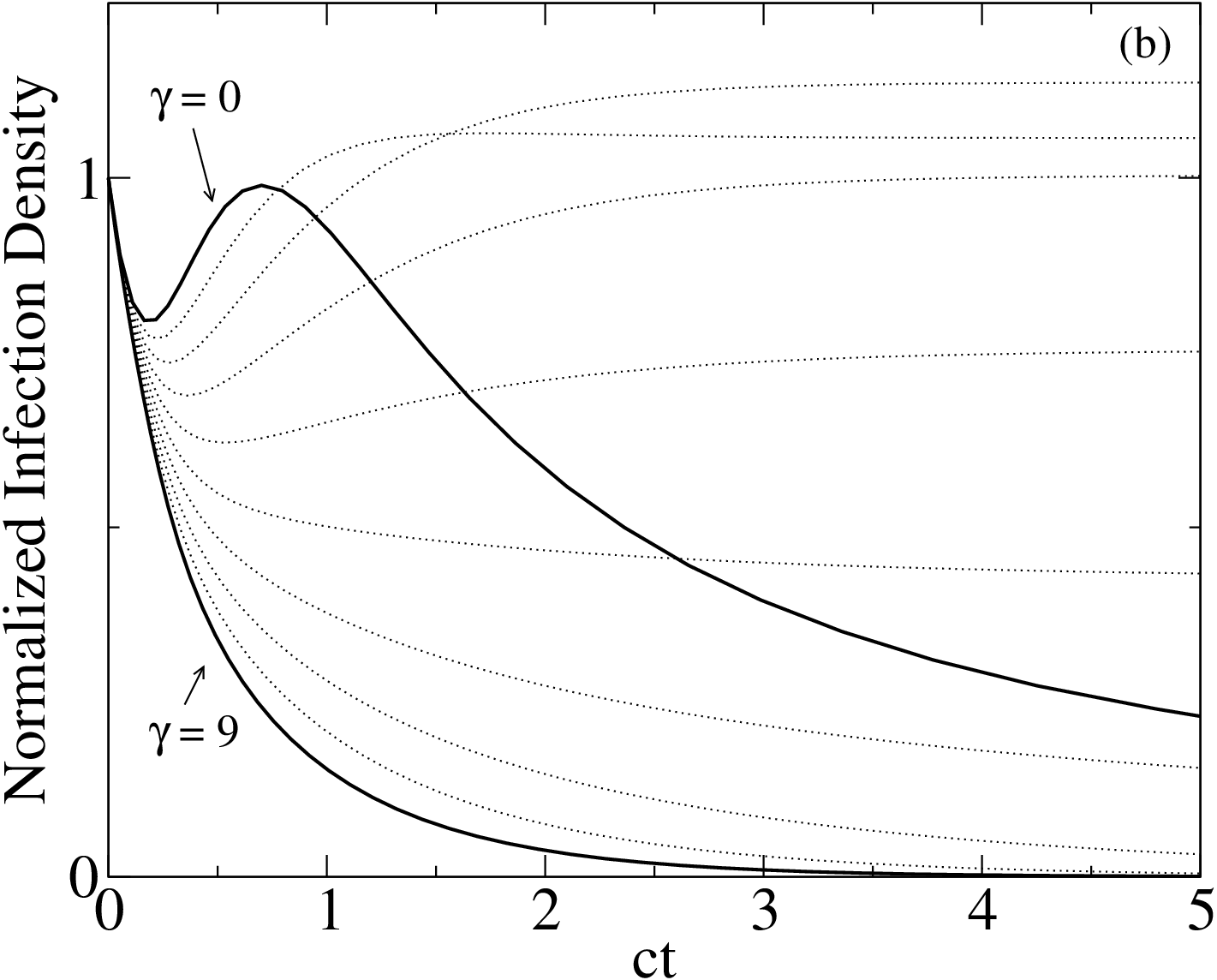}}}
\caption{Effect of variation in the growth rate on mice densities as
given by the mean field description. Steady state mice densities
are plotted in (a) against the
normalized growth rate $\gamma$ ($=g/c$) at which juveniles grow into
adults, and the evolution of the total density of infected mice,
normalized to its initial value,
is plotted in (b) against the normalized time
$ct$. Both exhibit non-monotonic behavior, for instance, a
rise in infection as $\gamma$ is increased from zero and a decay beyond
a certain value.
System parameters have been
chosen to be $\alpha_{0}=1.1$
and $\beta=15$. In (a),  the four mice densities have been shown
in the main figure and the total infected and susceptible densities, sums of adult and
juvenile contributions, are shown in the inset with axes identical to
the main figure. In (b), the ten curves are for $\gamma$ incremented by
1 from 0 to 9 and go down the graph in order for short times but for
long times approach equilibrium values that are not in the same order.
Notice also the peculiar non-monotonicity in time.}
\label{Fig1}
\end{figure}

With the help of the non-zero steady state values
$\overline{\mathcal{A}}$ and $%
\overline{\mathcal{B}}$ in Eq. (\ref{sumevol}) it is possible to obtain
(see Appendix) the steady state solutions of Eq. (\ref{normset}). The
system has the trivial solution ($\overline{\mathcal{A}}_{i}=\overline{%
\mathcal{A}}_{s}=\overline{\mathcal{B}}_{i}=\overline{\mathcal{B}}_{s}=0
$)
when $\beta <1$, and two possible steady states for $\beta >1$. They
represent respectively the non-infected and infected phase. The former
is
given by
\begin{eqnarray}
\overline{\mathcal{B}}_{i} &=&0,  \nonumber \\
\overline{\mathcal{B}}_{s} &=&\overline{\mathcal{B}},  \nonumber \\
\overline{\mathcal{A}}_{i} &=&0,  \nonumber \\
\overline{\mathcal{A}}_{s} &=&\overline{\mathcal{A}},  \label{ssnoninf}
\end{eqnarray}
while the latter is given by
\begin{eqnarray}
\overline{\mathcal{B}}_{i}
&=&\frac{\sqrt{\mathcal{F}^{2}+8\mathcal{E}}-%
\mathcal{F}}{4\alpha _{0}} ,  \nonumber \\
\overline{\mathcal{B}}_{s}
&=&\overline{\mathcal{B}}-\overline{\mathcal{B}}%
_{i},  \nonumber \\
\overline{\mathcal{A}}_{i} &=&\frac{\overline{\mathcal{B}}_{i}\left(
\gamma
+\alpha _{0}\overline{\mathcal{A}}\right) }{\alpha
_{0}\overline{\mathcal{B}}%
_{i}+1+\overline{\mathcal{A}}+\overline{\mathcal{B}}} ,  \nonumber \\
\overline{\mathcal{A}}_{s}
&=&\overline{\mathcal{A}}-\overline{\mathcal{A}}%
_{i},  \label{ssinf}
\end{eqnarray}
wherein $\mathcal{F}=2(\gamma +1)+\overline{\mathcal{A}}\left( 3+\alpha
_{0}\right) +\overline{\mathcal{B}}\left( 3-2\alpha _{0}\right) $ and $%
\mathcal{E}=\alpha _{0}\overline{\mathcal{B}}\left( \gamma +\alpha _{0}%
\overline{\mathcal{A}}\right) -\left[ \gamma +\overline{\mathcal{A}}+%
\overline{\mathcal{B}}\left( 1-2\alpha _{0}\right) \right] \left( 1+%
\overline{\mathcal{A}}+\overline{\mathcal{B}}\right) $. A stability
analysis
(see Appendix) shows that the solution sets represented by Eqs.
(\ref{ssnoninf}) and (\ref
{ssinf}) exchange their stability through a transcritical bifurcation
when the normalized infection rate is larger than a critical value
$\alpha
_{c}$. By determining when $\mathcal{E}=0$ we can calculate the critical
value as shown in the Appendix. Equivalently, by studying the non-normalized mean
field equations, we can calculate the critical environment parameter
$K_{c}$:
\begin{equation}
K_{c}=\frac{\gamma +2\left(
1+\overline{\mathcal{A}}+\overline{\mathcal{B}}%
\right) }{2a_{0}\overline{\mathcal{A}}}\left\{
\sqrt{1+\frac{4\overline{%
\mathcal{A}}\left( \gamma
+\overline{\mathcal{A}}+\overline{\mathcal{B}}%
\right) \left( 1+\overline{\mathcal{A}}+\overline{\mathcal{B}}\right)
}{%
\overline{\mathcal{B}}\left[ \gamma +2\left( 1+\overline{\mathcal{A}}+%
\overline{\mathcal{B}}\right) \right] ^{2}}}-1\right\}.  \label{kcrit}
\end{equation}
Here $\overline{\mathcal{A}}+\overline{\mathcal{B}}$ is the total
population
at steady state, i.e., the carrying capacity of the system normalized
to $cK$:
\begin{equation}
\overline{\mathcal{A}}+\overline{\mathcal{B}}=(\beta -1)\frac{\gamma }{%
1+\gamma }\frac{\sqrt{1+2\xi }-1}{\xi }.  \label{carcap}
\end{equation}
An infected phase exists for $K>K_c$.

\subsection{Simulation Description}

\label{sec:sd}

It is well-known \cite{CantrellCosnerbook} that spatial aspects, not
accessible to mean field theory,  are
(obviously) very important in epidemiology, generally in ecology, and
that
they require a description, capable of addressing spatial correlations.
One way
to treat spatial resolution is to adopt kinetic level approaches as in
the
AK model, while another, a more detailed way, is to adopt an approach
based
on the evolution of the full configuration states. We have elected to
choose
the latter in the present paper. Analytic solutions are typically
impossible
in such an approach except for oversimplified models. Hence we resort
to simulations as done previously \cite{aguirre} for the AK model.
\begin{figure}[ht]
\centering
\mbox{\subfigure{\label{Fig2a}\includegraphics[height=0.4\columnwidth,width=0.46\columnwidth]{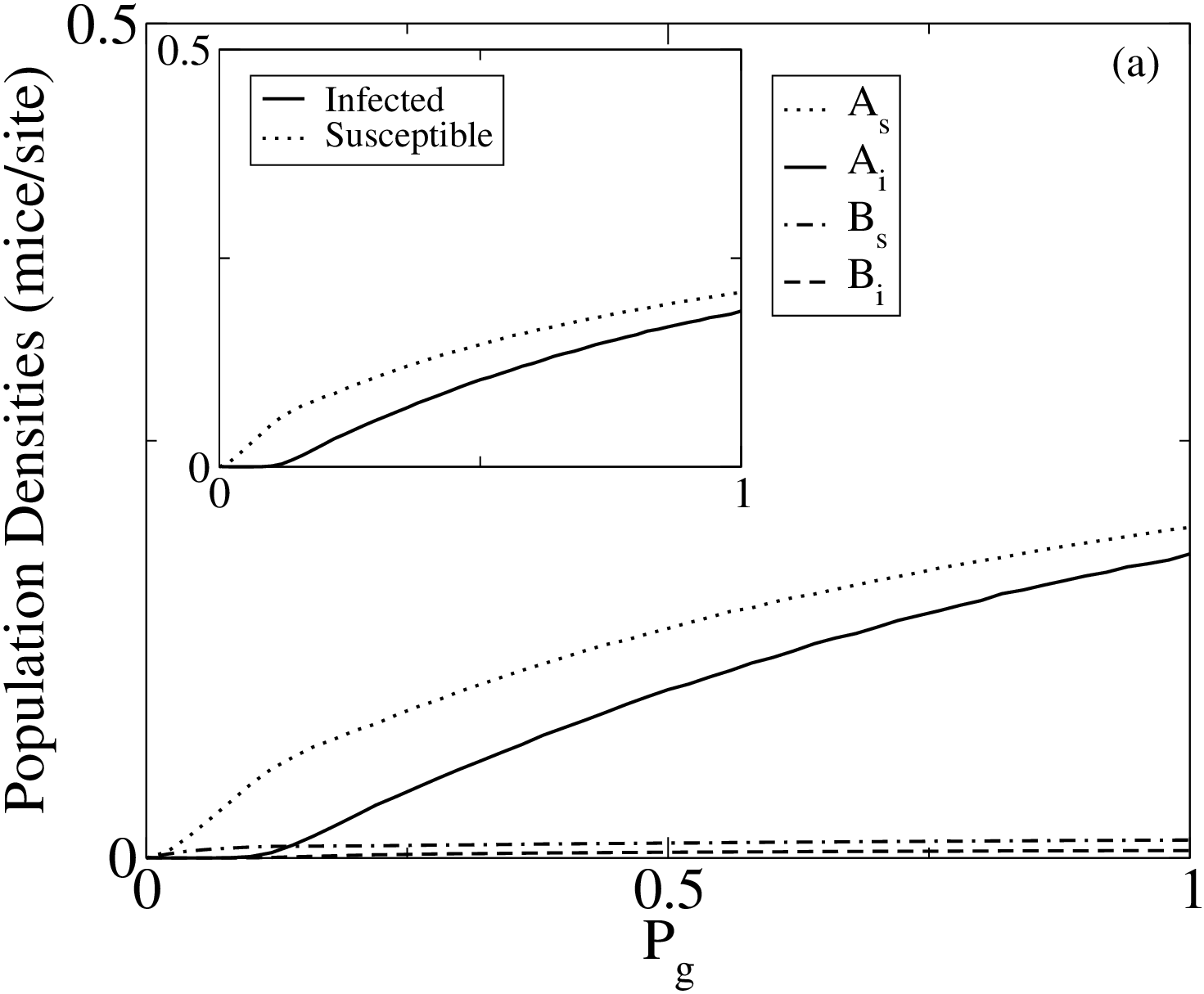}}
\hspace{0.03\columnwidth}
\subfigure{\label{Fig2b}\includegraphics[height=0.4\columnwidth,width=0.46\columnwidth]{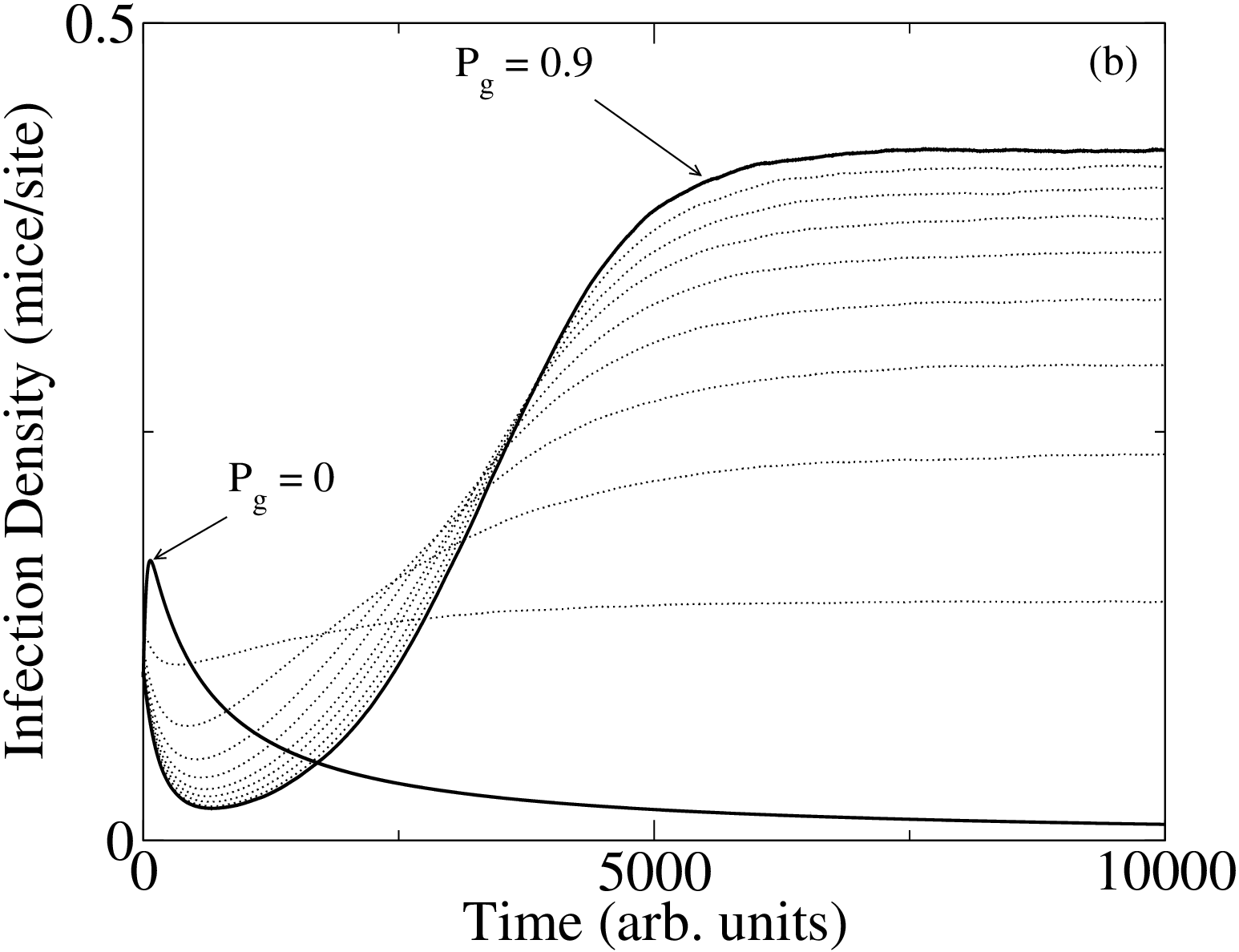}}}
\caption{Effect of variation in the growth probability as given by the
simulation description. Compare with Fig. \ref{Fig1}. Steady state mice
densities are plotted in (a) against $P_{g}$ and the time
evolution is shown in (b) where the ten curves correspond to increments of 0.1 in $P_g$.
Contrary to the mean-field model (see Fig.
\ref{Fig1}), where infected population as function of
the growth rate $\gamma$ always displays a non-monotonic behavior, here
the
infected population has a monotonic increase with
$P_{g}$. The parameters for both (a) and (b) are $P_{c}=0.01$, $%
P_{b}=0.011$, $P_{a}=0.3$ and $P_{K}=0.99$.}
\label{Fig2}
\end{figure}

Our simulations are carried out on a $L\times L$ square lattice with
each
site of the lattice corresponding to a small region in the landscape.
Moderately large lattices (with a total of $2^{14}$ sites) have been
used in
the simulations. The four subclasses of mice, (adults and juveniles in
susceptible and infected states) change their numbers, as time evolves,
in
accordance with rules which represent the model under consideration. At
each
time step, the juveniles may move but the adults not, the probability
for
the diffusive (random walk) motion being 0.125 for any of the eight
directions
of the square lattice. We consider the time step scaled to the diffusion
constant in this manner. An adult, infected or susceptible, gives birth
to a
susceptible juvenile with probability $P_{b}$. An adult dies by aging
with
probability $P_{c}$. If two or more mice meet at a site, one of them
may die
with probability $1-P_{K}$. If a susceptible mouse occupies the same
site as
an infected mouse, the former has probability $P_{a}$ of getting
infected in
the next time step. And if a juvenile mouse finds itself at a site
without
an adult, it grows up and settles at that site with probability
$P_{g}$. These
rules represent a simplified version of Eq. (\ref{lsmodel}) augmented
to the
configuration level. It is simplified in that the regions (`green
pastures')
where the juveniles may settle down have not been marked but the
process has
been represented through a probability of conversion. We have carried
out
full scale simulations which take into account the spatial extent of the
green pasture regions and will discuss the results elsewhere.

The simulation description thus has the parameters, $P_{b}$, $P_{c}$,
$P_{g}$%
, $P_{a}$, and $P_{K}$, in correspondence to the respective rates $b$,
$c$, $%
g$, $a_{0}$, and $K$ of the mean field equations analyzed above.
However,
the correspondence is not straightforward in all cases, as expected.
Extensive computer simulations were performed. We found that the system
reaches a steady state, after a transient. The main quantities we
analyzed
were the densities of the four populations of mice and the existence or
not
of infection in the steady-state. Densities are defined as the total
number
of mice in the lattice divided by the total volume (or area) of the
lattice.
Since the existence of more than one adult per site is not allowed, the
density of adults lies between 0 and 1. By contrast, the juvenile
density
may exceed 1.

\section{Results}

\label{results}

The main difference between the mean field and the simulation results
lies
in the effects of the quantities that govern the growth of juveniles
into
adults, rate $g$ and probability $P_{g}$, respectively. We show this
dependence respectively in Figs. \ref{Fig1} and \ref{Fig2}. From Fig.
\ref{Fig1a} we see a non-monotonic dependence of the steady state
infected
population densities as $\gamma $ increases. This non-monotonicity is
also
evident in Fig. \ref{Fig1b} where the time evolution for the entire
infected
population is shown for different values of $\gamma $. In the limit
$\gamma
\rightarrow 0$ in Eq. (\ref{sumss}) $\overline{\mathcal{A}},\overline{%
\mathcal{B}}\rightarrow 0$ since no new adult can be `born' (i.e.,
produced by conversion of a juvenile through growth), the juvenile
population cannnot be regenerated and dies out because of competition,
and
eventually no population can be sustained. The juveniles are
responsible for
spreading the infection to the adults and a larger number of them tends
to
increase the infected population. However if the juveniles convert into
adults too fast, there are less mobile carriers of infection, the
number of infected juveniles decreases (note that
juveniles are born susceptible and never infected, this being a
Hantavirus
characteristic), and eventually the infection disappears. In other
words we
can say that as $\gamma $ grows, the critical environment parameter
$K_{c}$
eventually becomes smaller than the system $K$.
\begin{figure}[ht]
\centering
\mbox{\subfigure{\label{Fig3a}\includegraphics[height=0.4\columnwidth,width=0.46\columnwidth]{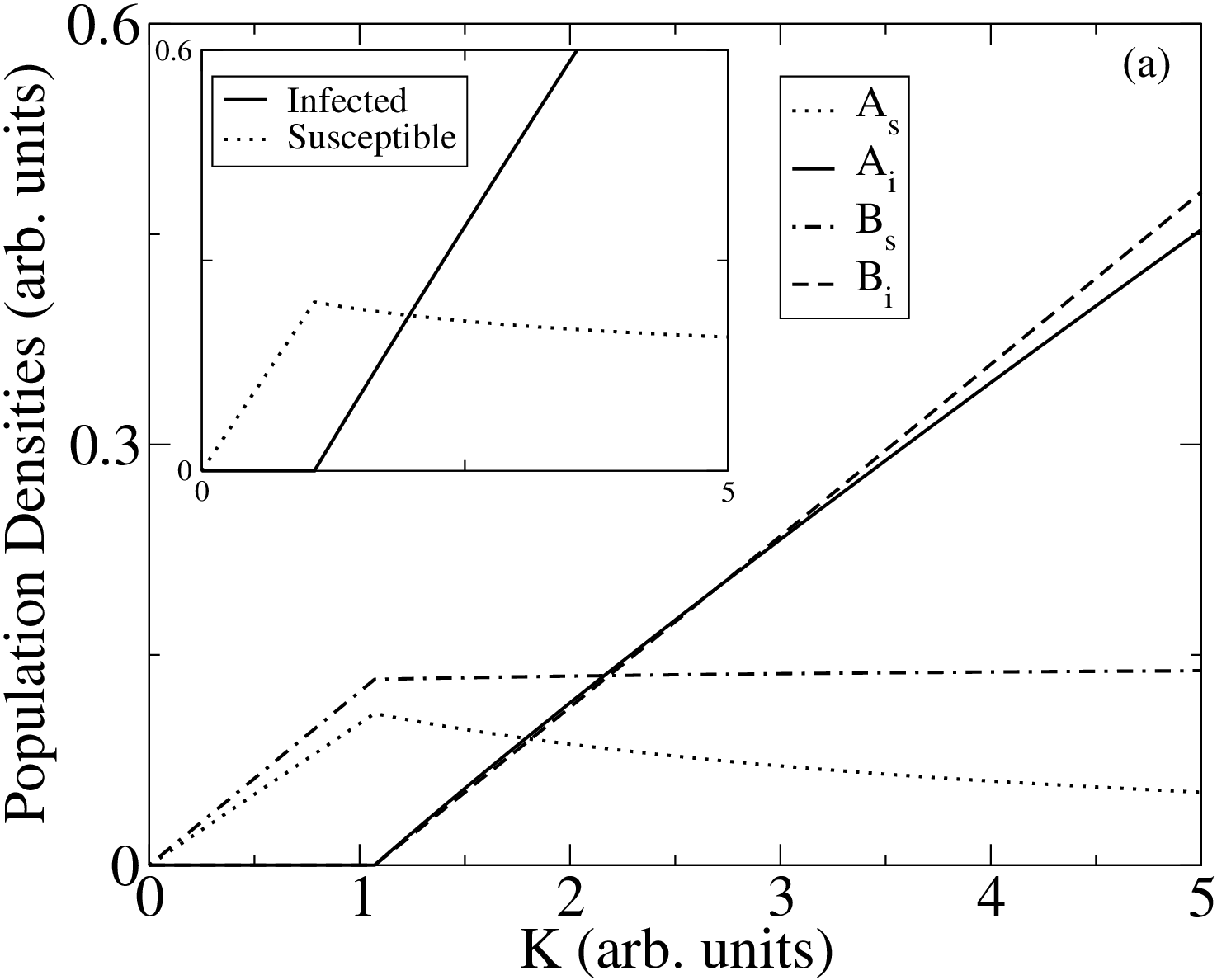}}
\hspace{0.03\columnwidth}
\subfigure{\label{Fig3b}\includegraphics[height=0.4\columnwidth,width=0.46\columnwidth]{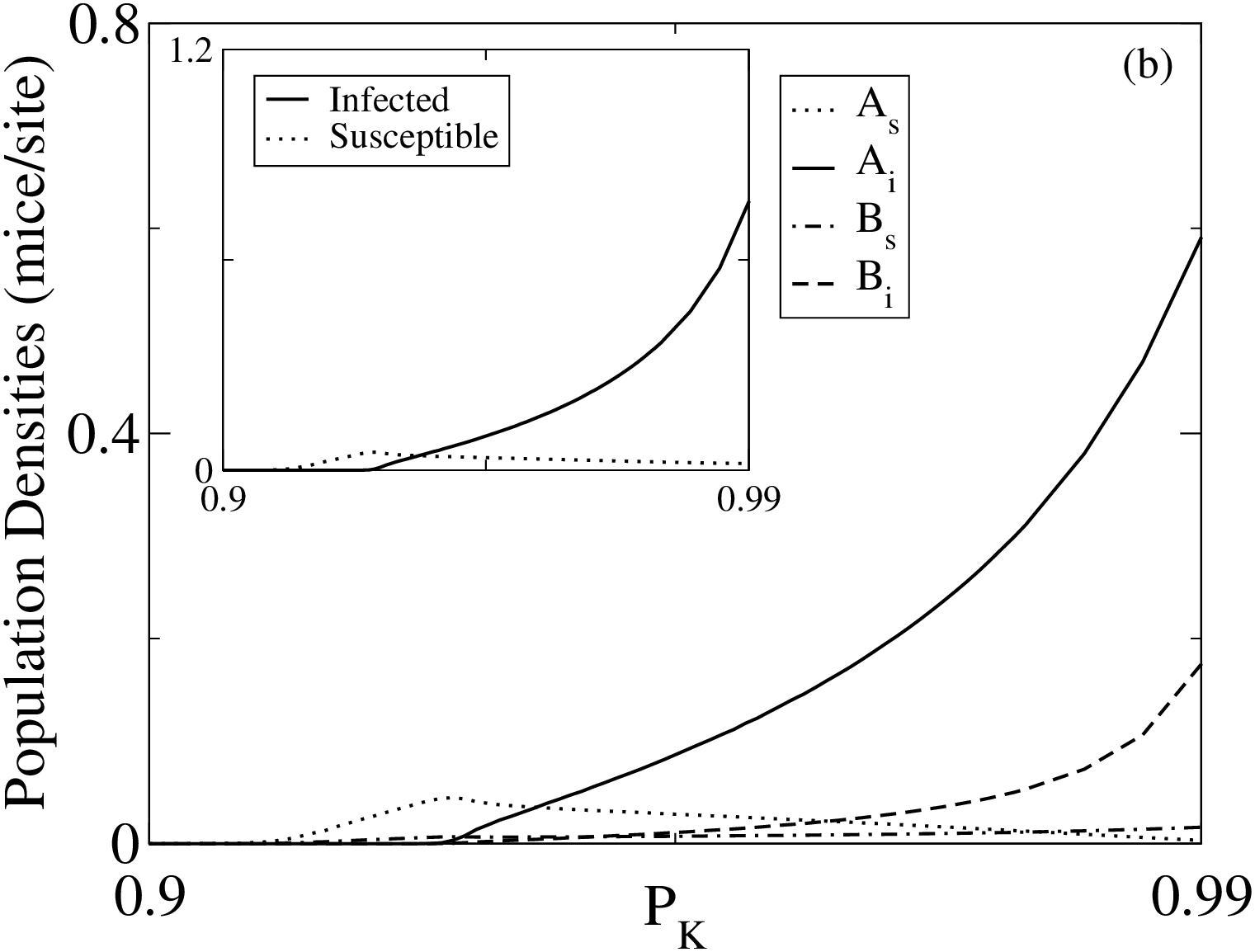}}}
\caption{Steady state densities versus the environment
parameter plotted from the mean field description in (a) and the
simulation description in (b). The x-axis coordinate in the inset and
the main figure is $K$ in (a) and $P_{K}$ in (b). Other parameters are
taken to be
$\beta=1.5$, $\gamma=1$ and $a_{0}=1$ in arbitrary units in (a) and
$P_{c}=0.01$, $P_{b}=0.014$, $P_{g}=0.1$
and $P_{a}=0.3$ in (b). The behavior is similar in both levels of
description. When the
environment parameter is large enough, an infected phase emerges. Once
that
happens, the increase of the population is due only to a larger density
of
infected animals, the susceptible population decreasing its overall
density to
a constant. The susceptible population in the simulation description
becomes zero for $P_{K}$ smaller than a critical value that
depends primarily on the value of $P_{b}$ relative to $P_{c}$. If $P_{b}$ is sufficiently
close to $P_{c}$, as is the case shown here, competition and adult death processes
drive the entire system to extinction at a nonzero value of $P_{K}$.}
\label{Fig3}
\end{figure}

In Fig. \ref{Fig2a} we show the $P_{g}$ dependence of the steady state
populations in the simulation description. This is a noteworthy
difference
from Fig. \ref{Fig1a}: a monotonic increase of the infected population.
This
difference has to be ascribed to a spatial correlation effect. Since
there
can be only one adult per site, an increase in $P_{g}$ has also the
effect
of reducing the number of available sites for the juveniles. For
sufficiently large values of $P_{g}$, clusters of adults start to form,
creating confined regions in the landscape where the juveniles are
constrained to roam.
This in turn increases dramatically the average time necessary for a
juvenile to find an available site to settle down. Unfavorable effects
of the growth of juveniles into adults observed in the mean field
theory do not therefore occur in the simulations as a result of the
spatial correlations set up by the cluster formation in the adults.
The monotonic increase of the infected
population as function of $P_{g}$ can also be observed in Fig.
\ref{Fig2b}
where the time evolution for different $P_{g}$ values is depicted. In
order to verify the validity of the above explanation of the difference
in the mean field and simulation predictions, we studied a
\emph{pseudo-model},
intermediate between the two descriptions. In the pseudo-model, a
juvenile can move to an arbitrary position and not only to a
nearest-neighbor site, thus being able to jump the barriers set up by
the clusters formed in the adult population. Careful simulations we
have carried out show that, indeed, the steady state
populations have an infected phase that decays to zero beyond a
critical value of $P_{g}$: the pseudo-model predicts the same
qualitative behavior as the mean field theory.

The struggle for resources described at the mean field level by the
environment parameter $K$ is here represented by the probability of
survival $P_{K}$: the probability of dying via competition for resources
is $%
1-P_{K}$. In Fig. \ref{Fig3a}  we show the steady state
populations as function of $K$ in the mean field description and in
Fig. \ref{Fig3b} the corresponding results as function of $P_{K}$) in
the simulation description. The behaviour in the two cases is similar:
beyond the critical value of $K$ ($P_{K}$) the infected population
increases with a
sublinear dependence on $K$ ($P_{K}$).

\section{Concluding remarks}

\label{conclusions}

From a perspective of concepts that have been successful in physics, our
present model of the mice assembly may be described by the term \emph{%
liquid-solid} because it describes a class of mice that move freely as
do
the molecules of a liquid, and another class of mice that move in the
neighborhood of fixed positions (the burrows) as do the molecules of a
solid, which vibrate around fixed lattice sites. Therefore, for the
purposes
of the following discussion, we will call our model the liquid-solid
(LS)
model.

Although the LS model was constructed by generalizing the structure of
the
AK model, the two have really only one parameter truly in common: the
environment parameter $K$. The birth rate $b$ in the AK model describes
the
emergence of new mice from all mice whereas it produces only juveniles
from
adults in the LS model. The death rate $c$ in the AK model is similarly
applicable to all mice but in the present version of the LS model it
applies
only to the adults. In addition to depletion via competition for resources,
the disappearance of the juveniles is
assumed to occur only through the growth rate $g$ when they grow up
into adults. The
diffusion constant $D$ describes the motion of all mice in the AK model
but
only of the juveniles in the LS model, the adults being stationary.
Therefore, we make comparison comments regarding the two models by
looking at how the infection depends on $K$.
For simplicity we discuss this comparison at the mean field level.

At first sight, it might
appear
that the LS model does not reduce to the AK counterpart in any
situation. However there exists one such limit. Let $K$ be infinite so
that the
carrying capacity is infinite also. To eliminate runaway solutions in
the
steady state, let us take the birth and death rates equal in the AK
model so
that $b_{AK}=c_{AK}$, and let us correspondingly take the birth-death ratio
$\beta
=1$ in the LS model. Consider now the limit in which the growth rate
$g$ of the
juveniles and the death rate $c$ of the adults in the LS model are much
larger
than the rate of infection between adults and juveniles but such that $%
\gamma=g/c$ is finite. In such a situation the adults constitute an
isolated
reservoir of the infection: they do not play any role in spreading it
since
an adult is infected only if it `grew up' from an infected juvenile. In
particular, it can be shown that by considering $a_{1}=a_{AK}$ and
$\gamma
=\left[ \left( 1+4c_{AK}\right) ^{1/2}+1\right] /2$, where $a_{AK}$ is
the
infection parameter in the AK model, our present model at steady state
gives
\emph{exactly} the same analytic dependence of the AK model for the
infected and
susceptible population as function of $c_{AK}$ and $a_{AK}$.

\begin{figure}[ht]
\centering
\mbox{\subfigure{\label{Fig4a}\includegraphics[height=0.4\columnwidth,width=0.46\columnwidth]{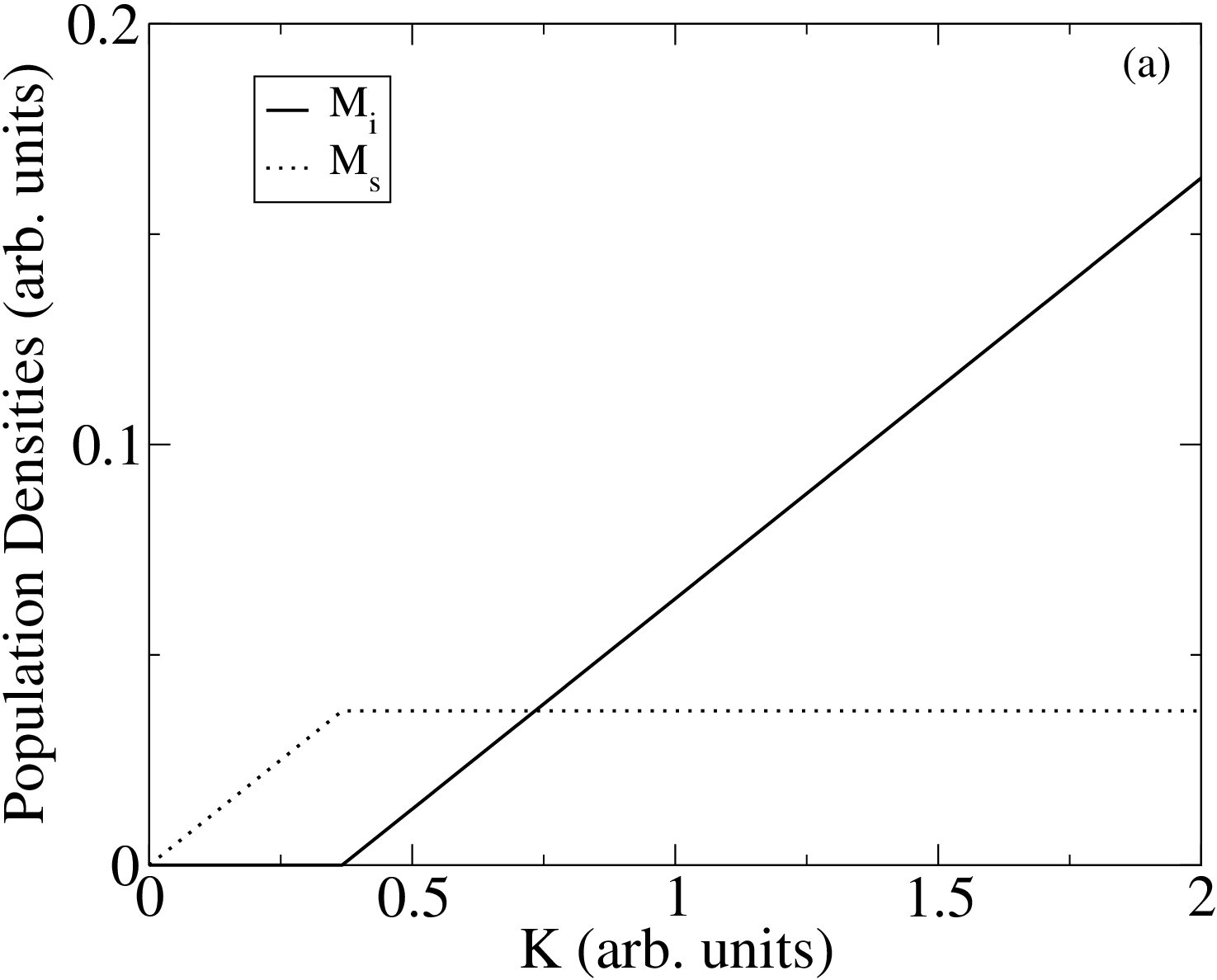}}
\hspace{0.03\columnwidth}
\subfigure{\label{Fig4b}\includegraphics[height=0.4\columnwidth,width=0.46\columnwidth]{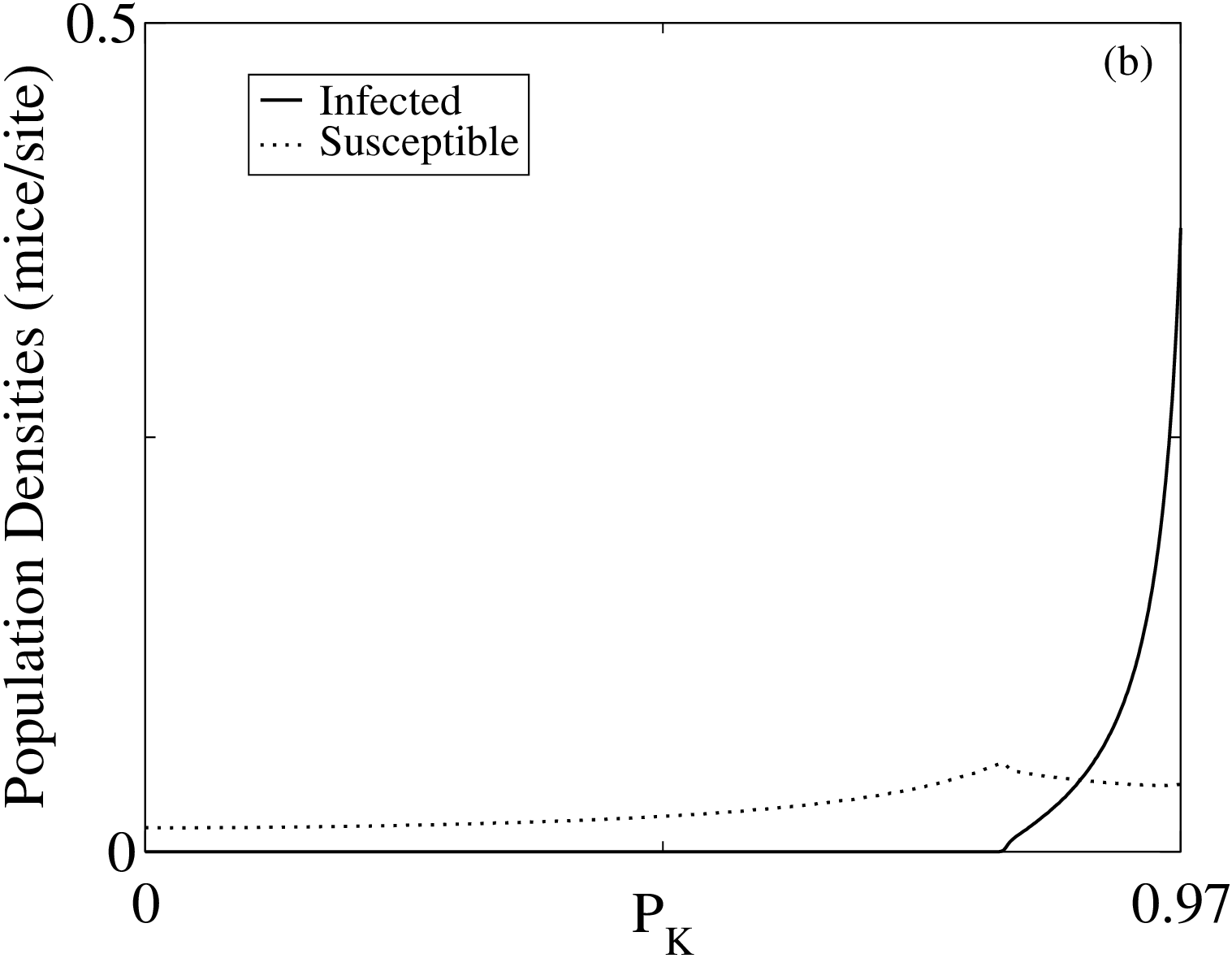}}}
\caption{Steady state densities of the Ak model as function of the environment
parameter plotted from the mean field description in (a) and the
simulation description in (b). The x-axis is the environment parameter
$K$ in (a) and the probability $P_{K}$ in (b).
Other parameters are
taken to be
$\beta=1.1$, and $a_{AK}=30$ in arbitrary units in (a) and
$P_{c}=0.01$, $P_{b}=0.02$,
and $P_{a}=0.6$ in (b).}
\label{Fig4}
\end{figure}

This equivalence suggests a simple way to compare the two models, when
$K$ is
finite, by considering that no transmission of infection occurs between
adults and juveniles, i.e., $a_{0}$ is taken to be zero in the LS model.
In such a scenario, the two models have the same qualitative
dependence of the infected and susceptible populations as function of
the
environment parameter: for $K>K_{c}$ the susceptible population remains
constant while the infected population increases linearly. In other
words, by
increasing $K$, the additional population (proportional to $K$) that the environment can
sustain
eventually becomes infected at steady state. In Fig. \ref{Fig4} we show this dependence
for the AK model at the mean field (Fig. \ref{Fig4a}) and at the simulation description
(Fig. \ref{Fig4b}), respectively. The behavior is similar in both levels of
description. When the
environment parameter is large enough, an infected phase emerges. Once
that
happens, the increase of the population is due only to a larger density
of
infected animals, the susceptible population remaining constant.
By contrast, as previously
shown in Fig. \ref{Fig3}, the infected
populations increase while the suscpetible populations decrease to a non zero value
as $K>K_{c}$. At the mean field
description it is also evident that
the infected population increases only sub-linearly as function of $K$ and not linearly
as in the AK model.

We indicate how our theoretical predictions may be used in conjunction  with observations in the field to test the validity/applicability of the LS model. The qualitative differences between the AK and LS models discussed above could be exploited in analyzing
data from \emph{mark-recapture} observations. In such observations, traps are set up in regions in the landscape and mice are caught, examined, marked, and released. Information about a variety of features is gathered including infection status and age. Juveniles have clear physionomical characteristics
that distinguish them from the adults \cite{stickel}. If all other effects are controlled,
supplementing additional food homogeneously over the terrain would be a way
to increase the environment parameter $K$. Experiments that exploit this feature
may allow one to determine if, in the presence of infection, the susceptible population
remains constant or decreases as the amount of food is increased. This would allow us
to establish whether the juveniles are indeed the main carrier of the disease and
whether augmenting the AK considerations
to the LS model is the correct way of analyzing the spread of the Hantavirus infection.

Since in both models the value of $K_{c}$ depends on all the other
parameters, there are obviously situations in which one model predicts
infection while the other does not. However, if we compare the amount of
infection as $K$ increases beyond $K_{c}$, our analysis shows that the
AK
model gives more infection than the LS model. Surely, this is to be
expected since, in the latter, part of the
population (adults) is stationary and transmit infection less
efficiently. This is an important consequence of the existence of home
ranges determined quantitatively in our recent work
\cite{panama,sevilleta,jtb}. Our analysis allows us to quantify that
consequence, i.e. to determine how much the reduced motion of the
adults
decreases the transmission of infection compared to the AK model.
While, for simplicity, we have
considered here the extreme limit of zero overlap between neighboring
home
ranges, to what extent the degree of home range overlap will change our
conclusions is
an open problem and the subject of our current investigations.

To keep our analysis focussed on essentials, we have considered in the
present paper the case in which the juvenile population is depleted
only through growth into adults and not through death. We have carried
out mean field as well as spatially resolved studies of the situation
when this simplification does not apply. This is important because, in
their itinerant attempts to find suitable places for their own home
ranges, juveniles are surely exposed to predators that kill them.
Details of the investigation will be given elsewhere \cite{ana} but we
report here that we have uncovered the epidemiologically noteworthy
possibility of
buffering the transmission of the infection by introducing predators in
the landscape.  Sustained by a sufficiently large $K$ (environment
resources), a large \emph{susceptible}
population of mice can exist but with no infection as a consequence of
interaction with  predators.

Among additional avenues of theoretical research based on the LS model
are the study of spatial correlations based on modern techniques
\cite{hutt}, the extension of our analysis by specifying the spatial
dependence of rates $G(x)$ in Eq. (\ref{lsmodel}) so that the effects
of the location of the `green pastures' can be ascertained, and the
comparison of our predictions to observations in the field concerning
infection spread. The latter effort is particularly important because
it will allow us to explore the applicability and practical relevance
of home range inclusion in the theory of the spread of the Hantavirus.
Research in all these directions is under way.

\begin{acknowledgments}
We acknowledge valuable conversations with Bob Parmenter and Terry
Yates about
the Hantavirus epidemic and with Marcelo Kuperman about aspects
of its description. This work was supported in part by the NSF
under grant no.
INT-0336343, by NSF/NIH Ecology of Infectious Diseases under grant no.
EF-0326757, by
CONICET (PEI 6482), by ANPCyT (PICT-R 2002-87/2),  and by DARPA under
grant no.
DARPA-N00014-03-1-0900. We acknowledge the Center for High Performance
Computing, UNM, for making available to us their
computing resources.
\end{acknowledgments}

\section{Appendix}

\label{appendix}

Since the total adult and juvenile population in (\ref{sumss}) is known
in
terms of the parameters $\xi $ and $\gamma $, the set of equations (\ref
{normset}) at steady state can be simplified considerably to
\begin{eqnarray}
\alpha _{0}\overline{\mathcal{B}}_{s}\overline{\mathcal{A}}_{i}+\alpha
_{1}%
\overline{\mathcal{B}}_{s}\overline{\mathcal{B}}_{i}-
\overline{\mathcal{B}}%
_{i}\left( \gamma +\overline{\mathcal{A}}+\overline{\mathcal{B}}\right)
&=&0,  \nonumber \\
\overline{\mathcal{B}}_{i}+\overline{\mathcal{B}}_{s}
&=&\overline{\mathcal{B%
}}, \nonumber \\
(\gamma +\alpha
_{0}\overline{\mathcal{A}}_{s})\overline{\mathcal{B}}_{i}-%
\overline{\mathcal{A}}_{i}\left(
1+\overline{\mathcal{A}}+\overline{\mathcal{%
B}}\right)  &=&0,  \nonumber \\
\overline{\mathcal{A}}_{i}+\overline{\mathcal{A}}_{s}
&=&\overline{\mathcal{A%
}}. \label{setss}
\end{eqnarray}
The system (\ref{setss}) has only three possible solutions with the four
variables larger or equal to zero. The trivial one implies $\overline{%
\mathcal{A}}=\overline{\mathcal{B}}=0$. The other two solutions can be
obtained by reducing through substitution Eq. (\ref{setss}) to the
following
polynomial equation in $\overline{\mathcal{B}}_{i}$
\begin{eqnarray}
&&\overline{\mathcal{B}}_{i}\left\{ \alpha _{0}\alpha
_{1}\overline{\mathcal{%
B}}_{i}^{2}+\overline{\mathcal{B}}_{i}\left\{
\overline{\mathcal{A}}\left(
\alpha _{0}^{2}+\alpha _{0}+\alpha _{1}\right)
+\overline{\mathcal{B}}\left[
\alpha _{0}+\alpha _{1}\left( 1-\alpha _{0}\right) \right] +2\gamma
\alpha
_{0}+\alpha _{1}\right\} \right.   \nonumber \\
&&\left. +\left[ \gamma
+\overline{\mathcal{A}}+\overline{\mathcal{B}}\left(
1-\alpha _{1}\right) \right] \left( 1+\overline{\mathcal{A}}+\overline{%
\mathcal{B}}\right) -\alpha _{0}\overline{\mathcal{B}}\left( \gamma
+\alpha
_{0}\overline{\mathcal{A}}\right) \right\} =0,  \label{bieq}
\end{eqnarray}
from which it is easy to obtain the solutions shown in Eq.
(\ref{ssnoninf})
and (\ref{ssinf}). The study of the sign of
Eq. (%
\ref{bieq}) gives the condition for the existence of an infected phase
\begin{equation}
\alpha _{c}=\frac{\gamma +2\left( 1+\overline{\mathcal{A}}+\overline{%
\mathcal{B}}\right) }{2\overline{\mathcal{A}}}\left\{ \sqrt{1+\frac{4%
\overline{\mathcal{A}}\left( \gamma +\overline{\mathcal{A}}+\overline{%
\mathcal{B}}\right) \left(
1+\overline{\mathcal{A}}+\overline{\mathcal{B}}%
\right) }{\overline{\mathcal{B}}\left[ \gamma +2\left(
1+\overline{\mathcal{A%
}}+\overline{\mathcal{B}}\right) \right] ^{2}}}-1\right\} ,
\label{alphacrit}
\end{equation}
which can be converted to a $K_{c}$ as written in Eq. (\ref{kcrit}).
Notice
that the third root of the polynomial in (\ref{bieq}) can be shown
through a
numerical study to be always negative and it is thus discarded.

The stability analysis of the three solutions of (\ref{normset}) is
done by
calculating the Jacobian of the system at steady state $J\left(
\overline{%
\mathcal{A}}_{i},\overline{\mathcal{A}}_{s},\overline{\mathcal{B}}_{i},%
\overline{\mathcal{B}}_{s}\right) $

\begin{eqnarray}
J &=&\left(
\begin{array}{cc}
-\left(
1+\overline{\mathcal{A}}+\overline{\mathcal{B}}+\overline{\mathcal{A}%
}_{i}\right) & \alpha
_{0}\overline{\mathcal{B}}_{i}-\overline{\mathcal{A}}%
_{i} \\
-\overline{\mathcal{A}}_{s} & -\left(
1+\overline{\mathcal{A}}+\overline{%
\mathcal{B}}+\overline{\mathcal{A}}_{s}+\alpha
_{0}\overline{\mathcal{B}}
_{i}\right) \\
\alpha _{0}\overline{\mathcal{B}}_{s}-\overline{\mathcal{B}}_{i} & -%
\overline{\mathcal{B}}_{i} \\
\beta -(\alpha _{0}+1)\overline{\mathcal{B}}_{s} & \beta
-\overline{\mathcal{%
B}}_{s}
\end{array}
\right.  \nonumber \\
&&\left.
\begin{array}{cc}
\alpha _{0}\overline{\mathcal{A}}_{s}-\overline{\mathcal{A}}_{i} & -%
\overline{\mathcal{A}}_{i} \\
-\left( \alpha _{0}+1\right) \overline{\mathcal{A}}_{s} & \gamma
-\overline{%
\mathcal{A}}_{s} \\
2\alpha _{0}\overline{\mathcal{B}}_{s}-\left( \gamma
+\overline{\mathcal{A}}+%
\overline{\mathcal{B}}+\overline{\mathcal{B}}_{i}\right) & \alpha _{0}%
\overline{A}_{i}+(2\alpha _{0}-1)\overline{\mathcal{B}}_{i} \\
-(2\alpha _{0}+1)\overline{\mathcal{B}}_{s} & -\left[ \gamma
+\overline{%
\mathcal{A}}+\overline{\mathcal{B}}+\overline{\mathcal{B}}_{s}+\alpha
_{0}\left( \overline{\mathcal{A}}_{i}+2\overline{\mathcal{B}}_{i}\right)
\right]
\end{array}
\right)  \label{jacob}
\end{eqnarray}
The trivial solution has the following four eigenvalues
\begin{eqnarray}
\lambda _{1} &=&-1,  \nonumber \\
\lambda _{2} &=&-\gamma ,  \nonumber \\
\lambda _{3} &=&-\frac{\gamma +1}{2}\left( \sqrt{1+2\xi }+1\right) ,
\nonumber \\
\lambda _{4} &=&\frac{\gamma +1}{2}\left( \sqrt{1+2\xi }-1\right) ,
\label{eignullsol}
\end{eqnarray}
from which it is evident that $\lambda _{4}<0$ if $\beta <1$, while all
the
other eigenvalues are always negative. The trivial solution is thus
stable
if $\beta <1$ and it becomes unstable when $\beta >1$.

The polynomial characteristic $P(\lambda )$ of the Jacobian
(\ref{jacob})
associated with the solution (\ref{ssnoninf}) is equal to the product $%
P_{1}(\lambda )P_{2}(\lambda )$ where
\begin{eqnarray}
P_{1}(\lambda ) &=&\lambda ^{2}+\lambda \left[ 1+3\left(
\overline{\mathcal{A%
}}+\overline{\mathcal{B}}\right) +\gamma \right]  \nonumber \\
&&+\left( 1+2\overline{\mathcal{A}}+\overline{\mathcal{B}}\right) \left(
\overline{\mathcal{A}}+2\overline{\mathcal{B}}+\gamma \right) -\left(
\overline{\mathcal{B}}-\beta \right) \left(
\overline{\mathcal{A}}-\gamma
\right) ,  \label{polchar1}
\end{eqnarray}
and
\begin{eqnarray}
P_{2}(\lambda ) &=&\lambda ^{2}+\lambda \left[ 1+2\left(
\overline{\mathcal{A%
}}+\overline{\mathcal{B}}\right) -2\alpha
_{0}\overline{\mathcal{B}}+\gamma
\right]  \nonumber \\
&&+\left( 1+\overline{\mathcal{A}}+\overline{\mathcal{B}}\right) \left(
\overline{\mathcal{A}}+\overline{\mathcal{B}}-2\alpha
_{0}\overline{\mathcal{
B}}+\gamma \right) -\alpha
_{0}^{2}\overline{\mathcal{A}}\overline{\mathcal{B%
}}.  \label{polchar2}
\end{eqnarray}
The eigenvalues associated to $P_{1}(\lambda )$ are given by
\begin{equation}
\lambda _{1_{\pm }}=\frac{1+\gamma }{4}\left[ 1-3\sqrt{1+2\xi }\pm
\sqrt{%
2\left( 1+\xi +\sqrt{1+2\xi }\right) }\right] ,  \label{eigen1}
\end{equation}
with $\lambda _{1_{-}}<0$ for any $\beta $ and $\gamma $ and $\lambda
_{1_{+}}>0$ when $\beta <1$. The eigenvalues associated to
$P_{2}(\lambda )$
are negative for $\alpha _{0}<\alpha _{c}$ and they become positive
when $%
\alpha _{0}>\alpha _{c}$. The solution defined in Eq. (\ref{ssnoninf})
is
thus unstable if either $\beta <1$ or if $\alpha _{0}>\alpha _{c}$.

The study of the sign of the eigenvalues associated with the third
possible
steady state is done numerically and it is possible to show that
Eq. (%
\ref{ssinf}) represents a stable steady state if $\beta >1$ and $\alpha
_{0}>\alpha _{c}$ and become unstable if either $\beta <1$ or $\alpha
_{0}<\alpha _{c}$. The mean field description from the dynamical point
of
view has thus two transcritical bifurcations: one when the growth rate
$b$
equals the death rate $c$ and the other one when the infection rate
$\alpha
_{0}$ equals the critical infection rate $\alpha _{c}$ or similarly
when the
environment parameter $K$ equals the critical environment parameter
$K_{c}$.

\end{document}